\documentclass[english]{revtex4}
\usepackage[T1]{fontenc}
\usepackage{esint}

\makeatletter

\providecommand{\tabularnewline}{\\}

\usepackage{babel}
\makeatother

\usepackage{graphicx}

\begin{document}

\title{String Tension Scaling in High-Temperature Confined $SU(N)$ Gauge Theories }

\author{Peter N. Meisinger}
\email{pnm@physics.wustl.edu}
\author{Michael C. Ogilvie}
\email{mco@wuphys.wustl.edu}
\affiliation{%
Department of Physics, Washington University, St.\ Louis, MO 63130, USA
}%

\date{\today}

\begin{abstract}
$SU(N)$ gauge theories, extended with adjoint fermions having periodic
boundary conditions, are confining at high temperature for sufficiently
light fermion mass $m$. In the high temperature confining region,
the one-loop effective potential for Polyakov loops has a $Z(N)$-symmetric
confining minimum. String tensions associated with Polyakov loops
are calculable in perturbation theory, and display a novel scaling
behavior in which higher representations have smaller string tensions
than the fundamental representation.
In the magnetic sector, the Polyakov
loop plays a role similar to a Higgs field, leading to an apparent breaking
of $SU(N)$ to $U(1)^{N-1}$. This is turn yields a dual effective
theory where magnetic monopoles give rise to string tensions for spatial
Wilson loops. The spatial string tensions are calculable semiclassically
from kink solutions of the dual system.
We show that the spatial string tensions $\sigma^{(s)}_k$ associated
with each $N$-ality $k$ obey a variant of Casimir scaling
$\sigma^{(s)}_k /\sigma^{(s)}_1 \leq \sqrt{k(N-k)/(N-1)} $.
Although
lattice simulations indicate that the high temperature confining
region is smoothly connected to the confining region of low-temperature
pure $SU(N)$ gauge theory,
the electric and magnetic string tension scaling laws are
different and readily distinguishable.
\end{abstract}
\maketitle

\section{Introduction}

One of the long-standing problems of modern strong-interaction physics
is the origin of quark confinement \cite{Mandelstam:1974pi,'tHooft:1981ht,Greensite:2003bk}. 
Recently, results from theory and from
lattice simulations \cite{Unsal:2008ch,Myers:2007vc}
 have indicated the existence of a class of 
models based on $SU(N)$ gauge theories in which confinement can
be understood using semiclassical arguments.
The essential feature of this class of models is the deformation of the
underlying gauge theory in such a way that the effective potential
for the Polyakov loop is minimized in a confining phase
in which the Polyakov loop apparently breaks
$SU(N)$ to $U\left(1\right)^{N-1}$.
The form
of the Polyakov loop in this phase
in turn leads
to the existence of monopole solutions,
and an explicit connection between
monopoles and confinement along the lines
developed by Polyakov \cite{Polyakov:1976fu}.

In this class of models,
at least one Euclidean direction is taken to be compact, of length $L$.
As $L$ is decreased from infinity, the pure gauge theory
undergoes a deconfining phase transition at a
critical value of $L$.
Confinement is restored for small $L$ by a deformation
of the pure gauge action.
The simplest deformation which does this
is an addition to the gauge action
of terms non-local in the compact direction \cite{Myers:2007vc}.
Lattice simulations of this model
are consistent with a smooth connection of 
the high-temperature confining phase 
to the confining phase of the pure gauge theory.
A local action with similar properties
is obtained by adding Dirac fermions in the adjoint representation to the
gauge action, and it is this model we will consider here.
The novel feature of this model is that the adjoint fermions are given
periodic boundary conditions in the finite direction, rather than the
standard antiperiodic boundary conditions.
Recent lattice simulations of 
$SU(3)$ with periodic adjoint fermions \cite{Cossu:2009sq}
observe many of the same features seen
in \cite{Myers:2007vc}.
We will refer to the compact direction
as the timelike direction, writing $L$ as $\beta$,
suggesting a system at finite temperature $T=1/\beta$.
This is slightly misleading, because the transfer matrix
for evolution in the compact direction is not positive-definite.
Periodic
boundary conditions in the timelike direction imply that the generating
function of the ensemble, \emph{i.e.}, the partition function, is
given by\begin{equation}
Z=Tr\left[\left(-1\right)^{F}e^{-\beta H}\right]\end{equation}
where $F$ is the fermion number.
This graded ensemble, familiar from supersymmetry,
can be obtained from an ensemble $Tr\left[\exp\left(\beta\mu F-\beta H\right)\right]$
with chemical potential $\mu$ by the replacement $\beta\mu\rightarrow i\pi$.
This system can also be viewed as a gauge theory with periodic boundary
conditions in one compact spatial direction of length $L=\beta$. 
From this point of view, the transfer matrix is positive-definite,
and the partition function is dominated by the ground state energy.
We will use the language
of finite temperature gauge theory because of the well-developed relation
between $Z(N)$ symmetry breaking and the confinement-deconfinement
transition.

Finite temperature gauge theories
are advantageous in many respects for the study of confinement. The
Polyakov loop operator $P$, given by\begin{equation}
P\left(\vec{x}\right)=\mathcal{P}\exp\left[i\int_{0}^{\beta}dtA_{4}\left(\vec{x},t\right)\right]\end{equation}
 represents the insertion of a static quark into a thermal system
of gauge fields. It is the order parameter for the deconfinement phase
transition in pure $SU(N)$ gauge theories, with $\langle Tr_{F}P\rangle=0$
in the confined phase, and $\left< Tr_{F}P \right>\ne0$
in the deconfined phase. The deconfinement phase transition is associated
with the spontaneous breaking of a global $Z(N)$ symmetry $P\rightarrow zP$
where $z=\exp\left(2\pi i/N\right)$ is the generator of $Z(N)$.
The unbroken $Z(N)$ symmetry of the confined phase leads to a rich
set of conditions that expectation values must satisfy. The character
$Tr_{R}P$ of each irreducible representation of $SU(N)$ transforms
as $Tr_{R}P\rightarrow z^{k}Tr_{R}P$ under $P\rightarrow zP$ for
some $k\in\left\{ 0,..,N-1\right\} $. In the confined phase, the
expectation value $\left\langle Tr_{R}P\right\rangle $ is $0$ for
all representations that transform non-trivially under $Z(N)$, \emph{i.e.},
have $k\neq0$. For each representation $R$ with $k\neq0$, there
is a timelike string tension $\sigma_{R}^{(t)}$ associated with the
large-distance behavior of the correlation function\begin{equation}
\left<Tr_{R}P\left(\vec{x}\right)Tr_{R}P^{+}\left(\vec{y}\right)\right>\simeq\exp\left[-\frac{\sigma_{R}^{(t)}}{T}\left|\vec{x}-\vec{y}\right|\right].\end{equation}
On physical grounds, it is generally believed that all representations
with the same non-zero value of $k$ have the same string tension
$\sigma_{k}^{(t)}$. This timelike string tension $\sigma_{k}^{(t)}$
can be measured from the behavior of the correlation function \begin{equation}
\left<Tr_{F}P^{k}\left(\vec{x}\right)Tr_{F}P^{+k}\left(\vec{y}\right)\right>\simeq\exp\left[-\frac{\sigma_{k}^{(t)}}{T}\left|\vec{x}-\vec{y}\right|\right]\end{equation}
at sufficiently large distances, where $\sigma_{k}^{(t)}\left|\vec{x}-\vec{y}\right|$
represents the potential energy between widely separated groups of
$k$ quarks and $k$ antiquarks. Thus the study of confinement at
finite temperature using the Polyakov loop largely reduces to the
study of $Z(N)$ symmetry and the existence of a mass gap in Polyakov
loop two-point functions, as opposed to the area law for Wilson loops
appearing in confinement at zero temperature. 

The addition of adjoint representation fermions to $SU(N)$ gauge
theories preserves the global $Z(N)$ symmetry of the action.
With normal antiperiodic boundary conditions 
for the fermions, the perturbative
effective action for the Polyakov loop shows that the deconfined
phase is favored at high temperature.
As the adjoint fermion mass decreases from infinity,
the critical temperature of the deconfinement transition
decreases.
With periodic boundary conditions for the fermions, however,
this class of field theories
can avoid the transition to the deconfined phase found in the pure gauge
theory for sufficiently light fermion mass and high temperatures.
If the number of adjoint fermion
flavors $N_{f}$ is less than $11/2$, these systems are asymptotically
free at high temperature, and therefore the effective potential for
$P$ is calculable using perturbation theory. As shown below, the
system will lie in the confining phase at high temperature
 if the adjoint fermion mass
$m$ is sufficiently light and $1/2<N_{f}<11/2$. 
Evidence from lattice simulations indicates that the high-temperature
confining region is smoothly connected to the low-temperature confined
phase of the pure gauge theory \cite{Myers:2007vc},
indicating a continuous change in string tensions
from one region to the other.
In the high-temperature
confining region, electric string tensions can be calculated perturbatively
from the effective potential, and magnetic string tensions arise semiclassically
from non-Abelian magnetic monopoles. Thus the high-temperature confining
region provides a realization of one of the oldest ideas about the
origin of confinement.

The following section describes the calculation
of the effective potential, and demonstrates
the realization of confinement at high temperature. 
Section III derives the temporal string
tensions via a perturbative expansion around the confining minimum
of the effective potential. In section IV, we discuss spatial string
tensions as measured by spatial Wilson loops. As in the classic treatment
of the three-dimensional adjoint Higgs model by Polyakov \cite{Polyakov:1976fu}, 
monopoles are responsible for confinement. Following the recent work of Unsal
and Yaffe \cite{Unsal:2008ch}, we are able to calculate semiclassically the spatial string
tensions, which exhibit a new string tension scaling law. A final
section discusses our results and their possible confirmation by lattice
simulation. An appendix details a mathematical identity useful for 
calculating the effective potential in the confined region.

\section{High Temperature Confinement}

Consider a boson in a representation $R$ of the gauge group
with spin degeneracy s moving in a constant Polyakov loop background $P$.
The one-loop effective potential at
non-zero temperature and density is given by \cite{Gross:1980br,Weiss:1980rj}
 \begin{equation}
V_{b}=sT\int\frac{d^{d}k}{\left(2\pi\right)^{d}}Tr_{R}\left[\ln\left(1-Pe^{\beta\mu-\beta\omega_{k}}\right)+\ln\left(1-P^{+}e^{-\beta\mu-\beta\omega_{k}}\right)\right]\end{equation}
where periodic boundary conditions are assumed. With standard boundary conditions
(periodic for bosons, antiperiodic for fermions), 1-loop effects always
favor the deconfined phase. For the case of pure gauge theories
in four dimensions, the
one-loop effective potential can be written in the form
\begin{equation}
V_{gauge}\left(P,\beta\right)=\frac{-2}{\pi^{2}\beta^{4}}\sum_{n=1}^{\infty}\frac{Tr_{A}P^{n}}{n^{4}}\end{equation}
where the trace is over the adjoint representation.
This series is minimized, term by term if $P\in Z(N)$, so $Z(N)$
symmetry is spontaneously broken at high temperature. The same minima
are obtained for any bosonic field with periodic boundary conditions
or for fermions with antiperiodic boundary conditions. 

The use of periodic boundary conditions for the adjoint fermions dramatically
changes their contribution to the Polyakov loop effective potential.
In perturbation theory, the replacement $\beta\mu\rightarrow i\pi$
shifts the Matsubara frequencies from $\beta\omega_{n}=\left(2n+1\right)\pi$
to $\beta\omega_{n}=2n\pi$. The one loop effective potential is now
that of a bosonic field, but with an overall negative sign due to
fermi statistics \cite{Meisinger:2001fi}. The sum of the effective potential for the
fermions plus that of the gauge bosons gives\begin{equation}
V_{1-loop}\left(P,\beta,m,N_{f}\right)=\frac{1}{\pi^{2}\beta^{4}}\sum_{n=1}^{\infty}\frac{Tr_{A}P^{n}}{n^{2}}\left[2N_{f}\beta^{2}m^{2}K_{2}\left(n\beta m\right)-\frac{2}{n^{2}}\right].\end{equation}
Note that the first term in brackets, due to the fermions, is positive
for every value of $n$, while the second term, due to the gauge bosons,
is negative. 

The largest contribution to the effective potential at high temperatures
is typically from the $n=1$ term, which can be written simply as\begin{equation}
\frac{1}{\pi^{2}\beta^{4}}\left[2N_{f}\beta^{2}m^{2}K_{2}\left(\beta m\right)-2\right]\left[\left|Tr_{F}P\right|^{2}-1\right]\end{equation}
 where the overall sign depends only on $N_{f}$ and $\beta m$. If
$N_{f}\ge1$ and $\beta m$ is sufficiently small, this term will
favor $Tr_{F}P=0$. On the other hand, if $\beta m$ is sufficiently
large, a value of $P$ from the center, $Z(N)$, is preferred. Note
that an $\mathcal{N}=1$ super Yang-Mills theory would correspond
to $N_{f}=1/2$ and $m=0,$ giving a vanishing perturbative contribution
for all $n$ \cite{Davies:1999uw,Davies:2000nw}. 
In that case, non-perturbative effects lead to a confining effective potential
for all values of $\beta$.
In the case of $N_{f}\ge1$, each term in the effective potential will
change sign in succession as $m$ is lowered towards zero.
This suggests that it should be possible to obtain
a $Z(N)$ symmetric, confining phase at high temperatures using adjoint
fermions with periodic boundary conditions or some equivalent deformation
of the theory.

The existence of a $Z(N)$ symmetric, confining phase at high temperatures 
has been confirmed in $SU(3)$, where both lattice
simulations and perturbative calculations have been used to show that
a gauge theory action with an extra term of the form $\int d^{4}x\ a_{1}Tr_{A}P$
is confining for sufficiently large $a_{1}$ at arbitrarily high temperatures
\cite{Myers:2007vc}.
This simple, one-term deformation is sufficient for $SU(2)$ and
$SU(3)$. However, in the general case, a deformation with at least
$\left[\frac{N}{2}\right]$ terms is needed to assure confinement
for representations of all possible non-zero $k$-alities. Thus the
minimal necessary deformation is of the form
\begin{equation}
\sum_{k=1}^{\left[\frac{N}{2}\right]}a_{k}Tr_{A}P^{k}\end{equation}
 which is analyzed in detail in \cite{Myers:2008ey}. If all the coefficients
$a_{k}$ are sufficiently large and positive, the one-loop effective potential\begin{equation}
V_{1-loop}\left(P,\beta\right)=\frac{-2}{\pi^{2}\beta^{4}}\sum_{n=1}^{\infty}\frac{Tr_{A}P^{n}}{n^{4}}+\sum_{k=1}^{\left[\frac{N}{2}\right]}a_{k}Tr_{A}P^{k}\end{equation}
will be minimized by a unique set of Polyakov loop eigenvalues corresponding
to exact $Z(N)$ symmetry. 

The unique set of $SU(N)$ Polyakov eigenvalues invariant under Z(N)
is $\left\{ w,wz,wz^{2},..,wz^{N-1}\right\} $, where $z=e^{2\pi i/N}$
is the generator of $Z(N)$, and $w$ is a phase necessary to ensure
unitarity. We will order these eigenvalues in a matrix $P_{0}$ as
\begin{equation}
P_{0}=w\cdot diag\left[1,z,z^{2},..,z^{N-1}\right].\end{equation}
We can write $P_{0}$ in the form\begin{equation}
\left(P_{0}\right)_{jk}=\delta_{jk}\exp\left(i\phi_{0j}\right)\end{equation}
where $\phi_{0j}=\frac{\pi}{N}\left(2j-N-1\right)$ which represents
uniform spacing of the eigenvalues around the unit circle.
The matrix $P_{0}$ is gauge-equivalent to itself after a $Z(N)$
symmetry operation:\begin{equation}
zP_{0}=gP_{0}g^{+}.\end{equation}
This guarantees that $Tr_{F}\left[P_{0}^{k}\right]=0$ for any value
of $k$ not divisible by $N$, indicating confinement for all representations
transforming non-trivially under $Z(N)$ \cite{Meisinger:2001cq}.  

To prove that $P_{0}$ is a global minimum of the effective potential,
we use the high-temperature expansion for the one-loop effective potential
of a particle in an arbitrary background Polyakov loop gauge equivalent
to the matrix $P_{jk}=\delta_{jk}e^{i\phi_{j}}$. The one-loop effective
potential can be written as
\begin{equation}
V_{1-loop}=\sum_{j,k=1}^{N}(1-\frac{1}{N}\delta_{jk})\left[V_{B}\left(\phi_{j}-\phi_{k},0\right)-2N_{f}V_{B}\left(\phi_{j}-\phi_{k},m\right)\right]\end{equation}
 where $V_{B}\left(\theta,m\right)$ is given by \cite{Meisinger:2001fi}
 \begin{eqnarray}
V_{B}\left(\theta,m\right) & = & -\frac{m^{2}T^{2}}{\pi^{2}}\sum_{n=1}^{\infty}\frac{1}{n^{2}}K_{2}\left(n\beta m\right)\cos\left(\theta\right)\nonumber\\
 & = & -\frac{2T^{4}}{\pi^{2}}\left[\frac{\pi^{4}}{90}-\frac{1}{48\pi^{2}}\theta_{+}^{2}\left(2\pi-\theta_{+}\right)^{2}\right]+\frac{m^{2}T^{2}}{2\pi^{2}}\left[\frac{\pi^{2}}{6}-\frac{1}{4}\theta_{+}\left(2\pi-\theta_{+}\right)\right]\nonumber\\
 &  & -\frac{T^{4}}{2\pi}\sum_{l}'\left\{ \frac{1}{3}\left[\left(\beta m\right)^{2}+\left(\theta-2\pi l\right)^{2}\right]^{3/2}-\frac{1}{3}\left|\theta-2\pi l\right|^{3}-\frac{1}{2}\left|\theta-2\pi l\right|\beta^{2}m^{2}-\frac{\beta^{4}m^{4}}{16\pi\left|l\right|}\right\}\nonumber\\
 &  & -\frac{m^{4}}{16\pi^{2}}\left[\ln\left(\frac{\beta m}{4\pi}\right)+\gamma-\frac{3}{4}\right]\end{eqnarray}
and $\theta_{+}$ is $\theta$ shifted as necessary to lie between
$0$ and $2\pi$. The prime on the summation over $l$ indicates
that the term singular at $l=0$ is to be omitted. 
The $T^{4}$ term
dominates for $\beta m\ll1$, giving\begin{equation}
V_{1-loop}\approx\sum_{j,k=1}^{N}(1-\frac{1}{N}\delta_{jk})\frac{2\left(2N_{f}-1\right)T^{4}}{\pi^{2}}\left[\frac{\pi^{4}}{90}-\frac{1}{48\pi^{2}}\left(\phi_{j}-\phi_{k}\right)^{2}\left(\phi_{j}-\phi_{k}-2\pi\right)^{2}\right]\end{equation}
 which has $P_{0}$ as a minimum provided $N_{f}>1/2$. For any fixed
value of $N,$ a sufficiently small value of $\beta m$ will make
the confined phase stable. Even if the adjoint fermion mass is enhanced
by chiral symmetry breaking, as would be expected in a confining phase,
it should be of order $gT$ or less, and the higher terms in the expansion
of $V_{1-loop}$ can be neglected at sufficiently high temperature. 

It is interesting to study the stability of the high-temperature confined
region as a function of $N$. Myers and Ogilvie have
determined the phase diagram as a function of $\beta m$ and $N$
up to $N=20$
by numerically
minimizing $V_{1-loop}$ as a function of all the Polyakov
loop eigenvalues \cite{Myers:2008ey,Myers:2009df}. 
These models display a rich phase structure, in
which there may be many phase transitions. When $N$ is not prime,
there are typically phases where $Z(N)$ center symmetry is spontaneously
down to $Z\left(p\right)$ where $p$ is a factor of $N$. In the case
of $N$ even, the confined phase gives way to a phase with unbroken
$Z\left(N/2\right)$ symmetry as $\beta m$ increases. The phase with
unbroken $Z(N/2)$ is the ``least deconfined'' of the partially
confining phases. In the Appendix, we show that for $N$ even,
the minima of the effective potential
corresponding to the confined phase, which has $Z(N)$ symmetry,
and to a partially confined phase which has $Z(N/2)$ symmetry are
equal when
\begin{equation}
\left(2^{d+1}-1\right)V_{B}\left(\beta,0,0\right)=2N_{f}\left[2^{d+1}V_{B}\left(\beta,0,Nm/2\right)-V_{B}\left(\beta,0,Nm\right)\right].\end{equation}
A solution exists only for $N_{f}>1/2$
and $\left(\beta m\right)_{c}$ generally decreases as $N_{f}$ increases.
For the case $N_{f}=1$ and $d=4$ we find
\begin{equation}
\left(\beta m\right)_{c}\simeq\frac{4.00398}{N}
\end{equation}
which is consistent with the numerical results
of Myers and Ogilvie.
Thus, we see that for any given $T$ and $N$, there is a range of values of $m$ for which
the system is in the confined phase.

For $\beta m\gg1$, the one-loop potential favors the deconfined phase,
so there must be at least one phase transition as $\beta m$ is varied.
In the case of $SU(2)$, there is indeed a single phase transition
separating the confined and deconfined phases. Somewhat surprisingly,
the transition is first-order between the high-temperature confined
region and the deconfined phase. This behavior is different from the
Ising-like, second-order transition in the pure gauge theory 
\cite{Engels:1989fz,Bogolubsky:2004gi,Velytsky:2007gj},
but both behaviors are possible in spin systems with $Z(2)$ symmetry. 
Typically, a negative quartic term in a scalar potential stabilized
by higher-order terms leads to a tricritical point, where a 
line of second-order transitions intersects a first-order line.
It seems very likely that both transitions
are part of a critical line in the $m-T$ plane, and are separated
by a tricritical point. For $N\ge3$, the one-loop potential predicts
that one or more phases separate the deconfined phase from the confined
phase. In the case of $SU(3)$, a single new phase is predicted, and
has been observed in lattice simulations. For higher values of $N$,
there is a rich set of possible phases, including some where $Z(N)$
breaks down to a proper subgroup $Z(p)$. In such phases, particles
in the fundamental representation are confined, but bound states
of $N/p$ such particles are not \cite{Ogilvie:2007tj,Myers:2008ey}. 

\section{Temporal String Tensions}

The timelike string tension $\sigma_{k}^{(t)}$ between k quarks and
k antiquarks can be measured from the behavior of the correlation
function \begin{equation}
\left<Tr_{F}P^{k}\left(\vec{x}\right)Tr_{F}P^{+k}\left(\vec{y}\right)\right>\simeq\exp\left[-\frac{\sigma_{k}^{(t)}}{T}\left|\vec{x}-\vec{y}\right|\right]\end{equation}
at sufficiently large distances. Two widely-considered scaling behaviors
for string tensions are Casimir scaling, characterized by\begin{equation}
\sigma_{k}=\sigma_{1}\frac{k\left(N-k\right)}{N-1},\end{equation}
and sine-law scaling, given by\begin{equation}
\sigma_{k}=\sigma_{1}\frac{\sin\left[\pi k/N\right]}{\sin\left[\pi/N\right]}.\end{equation}
 For a more detailed discussion of string tension scaling laws, see \cite{Greensite:2003bk}. 

Timelike string tensions are calculable perturbatively in the high-temperature
confining region from small fluctuations about the confining minimum
of the effective potential \cite{Meisinger:2004pa}.
The one-loop three-dimensional effective
Lagrangian is\begin{equation}
\frac{T}{g^{2}}\sum_{j=1}^{N}\left(\nabla\phi_{j}\right)^{2}+\frac{T^{3}}{\pi^{2}}\sum_{n=1}^{\infty}\frac{\left|Tr_{F}P^{n}\right|^{2}-1}{n^{2}}\left[2N_{f}\beta^{2}m^{2}K_{2}\left(n\beta m\right)-\frac{2}{n^{2}}\right]\end{equation}
For small fluctuations, we write\begin{equation}
P=P_{0}e^{i\delta\phi}\end{equation}
 where $\delta\phi$ lies in the Cartan algebra of $SU(N)$, and therefore has $N-1$ independent components. The powers of the Wilson loop are \begin{equation}
Tr_{F}P^{k}=w^{k}\sum_{j=1}^{N}z^{jk}e^{ik\delta\phi_{j}}\end{equation}
 If $k$ is not divisible by $N,$ we have approximately\begin{equation}
Tr_{F}P^{k}\simeq ikw^{k}\sum_{j=1}^{N}z^{jk}\delta\phi_{j}=ikw^{k}\delta\tilde{\phi}_{k}\end{equation}
where $\delta\tilde{\phi}$ is the discrete Fourier transform of $\delta\phi$,
related by\begin{equation}
\delta\tilde{\phi}_{k}=\sum_{j=1}^{N}z^{jk}\delta\phi_{j}\end{equation}
\begin{equation}
\delta\phi_{j}=\frac{1}{N}\sum_{k=1}^{N}z^{-jk}\delta\tilde{\phi}_{k}.\end{equation}
 Note that the reality of $\delta\phi$ implies that $\delta\tilde{\phi}_{k}^{*}=\delta\tilde{\phi}_{N-k}$;
the last Fourier component, $\delta\tilde{\phi}_{N}$, is identically
zero, due the tracelessness of $\delta\phi$. If $k$ is divisible
by $N,$ we have instead\begin{equation}
Tr_{F}P^{k}\simeq N-\frac{1}{2}k^{2}\sum_{j=1}^{N}\left(\delta\phi_{j}\right)^{2}=N-\frac{1}{2N}k^{2}\sum_{m=1}^{N}\left(\delta\tilde{\phi}_{m}\delta\tilde{\phi}_{N-m}\right).\end{equation}

These formulae allow us to write the three-dimensional effective action
in terms of the $\delta\tilde{\phi}_{k}$ to quadratic order. For
each value of $k$, the terms in the potential which contribute have
$n\equiv k\, mod\, N$, $n\equiv N-k\, mod\, N$, or $n\equiv0\, mod\, N$.
We obtain a different {}``mass'' $\sigma_{k}^{(t)}/T$ for each
Fourier component $\delta\phi_{k}$. The string tensions are of order $g$:
\begin{eqnarray}
\left(\frac{\sigma_{k}^{(t)}}{T}\right)^{2}&=&g^{2}N\frac{2N_{f}m^{2}}{2\pi^{2}}\sum_{j=0}^{\infty}\left[K_{2}\left((k+jN)\beta m\right)+K_{2}\left((N-k+jN)\beta m\right)-2K_{2}\left((j+1)N\beta m\right)\right]\nonumber\\
&&-g^{2}N\frac{T^{2}}{3N^{2}}\left[3\csc^{2}\left(\frac{\pi k}{N}\right)-1\right]
\end{eqnarray}
where the gluon contribution has been summed in the last term. Note
that the symmetry $\sigma_{k}^{(t)}=\sigma_{N-k}^{(t)}$ is manifest
in this formula. The string tensions are continuous functions of $\beta m$.
The $m=0$ limit has the simple form 
\begin{equation}\label{eq:spatialtension}
\left(\frac{\sigma_{k}^{(t)}}{T}\right)^{2}=\frac{\left(2N_{f}-1\right)g^{2}T^{2}}{3N}\left[3\csc^{2}\left(\frac{\pi k}{N}\right)-1\right]\end{equation}
and is a good approximation for $\beta m\ll1$. This scaling law is
not at all like either Casimir or sine-law scaling, because the usual
hierarchy $\sigma_{k+1}^{(t)}\ge\sigma_{k}^{(t)}$ is here reversed.
Because we expect on the basis of $SU(3)$ simulations that the high-temperature
confining region is continuously connected to the conventional low-temperature
region, there must be an inversion of the string tension hierarchy
between the two regions for all $N\ge4$. For the case $N_{f}=1/2$
, corresponding to a single multiplet of adjoint Majorana fermions,
the perturbative string tension vanishes. As discussed in \cite{Davies:1999uw,Davies:2000nw},
it is the non-perturbative contribution to the effective potential
induced by monopoles that gives rise to the string tension in this
case. The large-$N$ limit of eqn. (\ref{eq:spatialtension})
 is smooth.
For fixed $k$ as $N\rightarrow\infty$, we have 
 \begin{equation}
\left(\frac{\sigma_{k}^{(t)}}{T}\right)^{2}\sim
\frac{\left(2N_{f}-1\right)\lambda T^{2}}{\pi^{2}k^{2}}\end{equation}
where $\lambda$ is the 't
Hooft coupling $g^{2}N$.

\section{Spatial String Tensions}

The confining minimum $P_{0}$ of the effective potential breaks $SU(N)$
to $U(1)^{N-1}$. This remaining unbroken Abelian gauge group naively
seems to preclude spatial confinement, in the sense of area law behavior
for spatial Wilson loops. However, as first discussed by Polyakov
in the case of an $SU(2)$ Higgs model in $2+1$ dimensional gauge
systems, instantons can lead to nonperturbative confinement \cite{Polyakov:1976fu}.
In the high-temperature confining region, the dynamics of the magnetic
sector are effectively three-dimensional due to dimensional reduction.
The Polyakov loop plays a role similar to an adjoint Higgs field,
with the important difference that $P$ lies in the gauge group,
while a Higgs field would lie in the gauge algebra. The standard topological
analysis \cite{Weinberg:1979zt} is therefore slightly
altered, and there are $N$ fundamental monopoles in the finite temperature
gauge theory \cite{Lee:1998vu,Kraan:1998kp,Lee:1998bb,Kraan:1998pm,Kraan:1998sn}
with charges proportional to the affine roots of $SU(N)$, given by
$2\pi\alpha_{j}/g$ where $\alpha_{j}=\hat{e}_{j}-\hat{e}_{j+1}$
for $j=1$ to $N-1$ and $\alpha_{N}=\hat{e}_{N}-\hat{e}_{1}$. Monopole
effects will be suppressed by powers of the Boltzmann factor $\exp\left[-E_{j}/T\right]$
where $E_{j}$ is the energy of a monopole associated with $\alpha_{j}$. 

In the high-temperature confining region, monopoles interact with
each other through both their long-ranged magnetic fields, and also
via a three-dimensional scalar interaction, mediated by $A_{4}$.
The scalar interaction is short-ranged, falling off with a mass of
order $gT$. The long-range properties of the magnetic sector may
be represented in a simple form by a generalized sine-Gordon model
which generates the grand canonical ensemble for the monopole/anti-monopole
gas \cite{Unsal:2008ch}. The action for this
model represents the Abelian dual form of the magnetic sector of the
$U(1)^{N-1}$ gauge theory. It is given by
\begin{equation}
S_{mag}=\int d^{3}x\left[\frac{T}{2}\left(\partial\rho\right)^{2}-2\xi\sum_{j=1}^{N}\cos\left(\frac{2\pi}{g}\alpha_{j}\cdot\rho\right)\right]\end{equation}
 where $\rho$ is the scalar field dual to the $U(1)^{N-1}$ magnetic
field. The monopole fugacity $\xi$ is given by $\exp\left[-E_{j}/T\right]$
times functional determinantal factors \cite{Zarembo:1995am}. 

This Lagrangian is a generalization of the one considered by Polyakov
for $SU(2)$, and the analysis of magnetic confinement follows along
the same lines \cite{Polyakov:1976fu}. The Lagrangian has $N$ degenerate inequivalent minima
$\rho_{0k}=g\mu_{k}$ where the $\mu_{k}$'s are the simple fundamental
weights, satisfying $\alpha_{j}\cdot\mu_{k}=\delta_{jk}$. In the
basis we use, $\mu_{k}$ is an $N$-dimensional vector of the form
$1/N\cdot\left\{ k,k,..,k-N,k-N\right\} $ with $\left(N-k\right)$
$k$'s and $k$ $\left(k-N\right)$'s. Note that $e^{2\pi i\mu_{k}}=z_{N}^{k}$
and the vector $\rho_{0N}$ is the zero vector. A spatial Wilson loop
in the $x-y$ plane
\begin{equation}
W\left[\mathcal{C}\right]=\mathcal{P}\exp\left[i\oint_{\mathcal{C}}dx_{j}\cdot A_{j}\right]\end{equation}
introduces a discontinuity in the $z$ direction
in the field dual to the magnetic field tensor.
We can move this discontinuity out to spatial infinity; then the string
tension of the spatial Wilson loop is the interfacial energy of a
kink $\rho(z)$ interpolating between the different vacua $\rho_{0k}$
as $z\rightarrow\pm\infty$ \cite{Unsal:2008ch}.

In the case of $SU(2)$, the Cartan algebra is one-dimensional, and
the kink solution is the sine-Gordon soliton. A BPS-type inequality
\cite{Bogomolny:1975de} 
gives the string tension as\begin{equation}
2\sqrt{8\xi T}\frac{g}{\pi}.\end{equation}
For higher $N$, the general kink solutions are not known analytically.
It is likely that the $k=1$ kink is given by a straight-line path
along the $\mu_{1}$ direction. It is easy to check that motion of
the Polyakov loop $Tr_{F}P\left(z\right)=\sum_{j=1}^{N}\exp\left(\frac{2\pi i}{g}\hat{e}_{j}\cdot\rho\right)$
is an arc in the complex plane from $N$ to $Nz_{N}$ along the boundary of allowed values
of $Tr_{F}P$ . In the case of $N=3$, we can prove this is the global
minimum. In this case the sum over affine roots can be written as\begin{equation}
-2\xi\sum_{j=1}^{3}\cos\left(\frac{2\pi}{g}\alpha_{j}\cdot\rho\right)=\xi\left[3-\left|\sum_{j=1}^{3}\exp\left(\frac{2\pi i}{g}\hat{e}_{j}\cdot\rho\right)\right|^{2}\right]\end{equation}
and an argument similar to the one given in \cite{Bhattacharya:1992qb}
shows that this solution is a global minimum.
There is a simple ansatz that generallzes this $k=1$ kink solution
to higher values of $k$. The ansatz is a straight line through the Lie algebra,
parametrized as $\rho(z) = g \mu_k q(z)$ \cite{Giovannangeli:2001bh}.
Figure 1 shows the $k=1,2,3$ paths in the $Tr_F P$ complex plane for $SU(6)$.
\begin{figure}
\centering
\includegraphics[width=.55\textwidth]{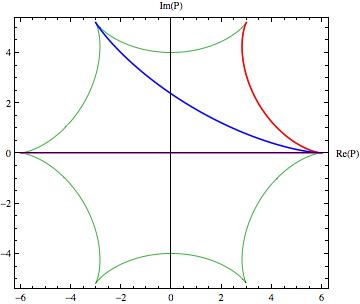}
\caption{$k=1,2,3$ paths in the  $Tr_F P$ complex plane for $SU(6)$. The $k=1$ path
is along the boundary of allowed values of $Tr_F P$}
\label{fig:1}
\end{figure}
For the general case, this straight line ansatz 
gives an action\begin{equation}
S_{mag}=\int d^{3}x\left[\frac{T}{2}g^{2}\frac{k\left(N-k\right)}{N}\left(\partial q\right)^{2}-2\xi\cos\left(2\pi q\right)\right]\end{equation}
yielding an upper bound of the form\begin{equation}
\sigma_{k}^{(s)}\le\frac{8}{\pi}\left[\frac{g^{2}T\xi}{N}k\left(N-k\right)\right]^{1/2}.\end{equation}
This bound is exact for $N=2$ or $3$ where there is only one independent
string tension. If, as seems likely, this result is exact for $k=1$
for all $N,$ this would give a useful bound on string tension ratios
of the form\begin{equation}
\frac{\sigma_{k}^{(s)}}{\sigma_{1}^{(s)}}\le\sqrt{\frac{k\left(N-k\right)}{N-1}}.\end{equation}
This square-root-Casimir behavior differs significantly from both
Casimir and sine-law scaling, and should be easily distinguishable
in lattice simulations. Table 1 compares the string tension scaling
laws we have found in the high temperature confined region
to Casimir and sine-Law scaling.

\begin{table}
\centering
\begin{tabular}{|c|c||c|c|c|c|}
\hline 
$N$ & $k$ & Casimir & sine-Law & $\sigma_{k}^{(t)}/\sigma_{1}^{(t)}$ & $\sigma_{k}^{(s)}/\sigma_{1}^{(s)}$\tabularnewline
\hline
\hline 
$4$ & $2$ & $1.33$ & $1.41$ & $0.40$ & $1.15$\tabularnewline
\hline 
$5$ & $2$ & $1.50$ & $1.62$ & $0.30$ & $1.22$\tabularnewline
\hline 
$6$ & $2$ & $1.60$ & $1.73$ & $0.27$ & $1.26$\tabularnewline
\hline 
$6$ & $3$ & $1.80$ & $2.00$ & $0.18$ & $1.34$\tabularnewline
\hline 
$7$ & $2$ & $1.67$ & $1.80$ & $0.26$ & $1.29$\tabularnewline
\hline 
$7$ & $3$ & $2.00$ & $2.25$ & $0.14$ & $1.41$\tabularnewline
\hline 
$8$ & $2$ & $1.71$ & $1.85$ & $0.26$ & $1.31$\tabularnewline
\hline 
$8$ & $3$ & $2.14$ & $2.41$ & $0.13$ & $1.46$\tabularnewline
\hline 
$8$ & $4$ & $2.29$ & $2.61$ & $0.10$ & $1.51$\tabularnewline
\hline
\end{tabular}
\caption{Comparison of Casimir and sine-Law scaling with temporal and spatial string tension scaling in the high-temperature confining region.}
\label{table:1}
\end{table}

\section{Conclusions}

We have been able to predict analytically a number of properties of
the high temperature confined region, which lattice simulations should
be able to confirm. For all values of $N$, there
exists a high-temperature confining region when the fermion mass is
sufficiently small.
The phase structure and thermodynamics of these
models are particularly rich for $N\ge4$. However, even in the case
of $SU(2)$, there is an interesting prediction of a first-order transition
between the deconfined phase and the high-temperature confined region
as the adjoint fermion mass is varied. It follows that there must
be a tricritical point in the $\beta-m$ plane somewhere on the critical
line separating the confined and deconfined phases, whose location
could be determined with lattice simulations.

We have shown that there is a perturbative
prediction for an inverted hierarchy of timelike string tensions,
beginning at $N=4$. This behavior is very different from
the string tension scaling of pure $SU(N)$ gauge theory
at zero temperature. 
It is very likely that the crossover between the two
behaviors is dramatic.
We have also developed
a semiclassical bound for spacelike string
tensions, which also indicates a deviation from
the string tension scaling of pure $SU(N)$.
However, the behavior of spacelike string tensions
in the high temperature confined region
is much closer to that expected in the pure gauge theory,
and the crossover between the two regions might be smooth.
It is intriguing that spacelike confinement is associated with
monopole condensation, the oldest scenario for quark confinement.
The predictions for both spacelike and timelike string tensions
are potentially testable in lattice simulations.

The large-$N$ behavior of the high-temperature confining region
is of obvious interest.
Unsal and Yaffe have argued \cite{Unsal:2008ch} that
the description of this
region as an $U(1)^{N-1}$ effective
theory is valid only when the inequality
$ T \gg N\Lambda $
is satisfied, where 
$\Lambda$ is the usual renormalization-group
invariant scale for an $SU(N)$ gauge theory.
If this inequality holds,
Abelian monopoles control
are responsible for area-law
behavior of spatial Wilson loops.
On the other hand, 
we have shown that temporal string tensions
and confinement,
as measured by temporal Polyakov loops,
require $T \gtrsim N m$.
In fact, these two inequalities may be
compatible, because of dynamical
generation of a fermion mass.

These predictions for the high-temperature region lead naturally to
additional questions that lattice simulations can address, but semiclassical
methods most likely cannot. Lattice simulations can explore the crossover
from conventional, low-temperature confining behavior to the behavior
predicted in the high-temperature confining region. Some features
can be studied in simulations where a simple deformation of the action
is used, as in \cite{Myers:2007vc}. In addition to
the string tensions, these features include monopole and instanton
densities, and the topological susceptibility. Other aspects will
require the inclusion of adjoint dynamical fermions in lattice simulations.
Chiral symmetry breaking is of particular interest. Unsal \cite{Unsal:2007vu,Unsal:2007jx} 
has proposed a detailed picture of chiral symmetry breaking which
can be independently checked by simulation. The accessibility of lattice
field configurations as well as conventional observables makes the
high-temperature confined region a natural place to explore the overlap
of theory and simulation. 

\appendix*
\section{}
The determination of the critical point between the confined phase,
with $Z(N)$ symmetry, and a phase with $Z(N/2)$ symmetry, is made
possible by a generalization of a Bernoulli polynomial summation formula
\cite{Gradshteyn:1994gr}. Let $f\left(\theta,M\right)$ have the form\begin{equation}
f_p\left(\theta,M\right)=\sum_{n=1}^{\infty}\frac{1}{n^{p}}g\left(nM\right)\left[e^{in\theta}+e^{-in\theta}\right]\end{equation}
 where $g$ is sufficiently well-behaved that the series converges.
Consider the finite sum\begin{equation}
\sum_{k=0}^{m-1}f_p\left(\theta+\frac{2\pi k}{m},M\right)=\sum_{n=1}^{\infty}\sum_{k=0}^{m-1}\frac{1}{n^{p}}g\left(nM\right)\left[e^{in\theta+2\pi ink/m}+e^{-in\theta-2\pi ink/m}\right].\end{equation}
The sum over $k$ gives a non-zero result only when $n$ is a multiple
of $m$, giving\begin{equation}
\sum_{k=0}^{m-1}f_p\left(\theta+\frac{2\pi k}{m},M\right)=m^{1-p}\sum_{n=1}^{\infty}\frac{1}{n^{p}}g\left(nmM\right)\left[e^{inm\theta}+e^{-inm\theta}\right]=m^{1-p}f_p\left(m\theta,mM\right).\end{equation}

This formula can be directly applied to the basic quantity for constructing
all one-loop effective potentials in $d$ spatial dimensions, the effective potential
of a massive charged boson in a constant $U(1)$ Polyakov loop background.
The $U(1)$ Polyakov loop is parametrized as 
$\exp\left(i\int_{0}^{\beta}dt\, A_{d}\right)=\exp\left(i\theta\right)$, 
and the standard periodic boundary conditions for bosons are used.
The finite-temperature component of the effective potential is\begin{equation}
V_{B}\left(\beta,\theta,M\right)=\frac{1}{\beta}\int\frac{d^{d}k}{\left(2\pi\right)^{d}}\ln\left[1-e^{-\beta\omega_{k}+i\theta}\right]+\frac{1}{\beta}\int\frac{d^{d}k}{\left(2\pi\right)^{d}}\ln\left[1-e^{-\beta\omega_{k}-i\theta}\right]\end{equation}
 where $\omega_{k}$ is the energy $\sqrt{k^{2}+M^{2}}$ \cite{Meisinger:2001fi}.
This can be expanded as\begin{equation}
V_{B}\left(\beta,\theta,M\right)=-2\left(\frac{M}{2\pi\beta}\right)^{\left(d+1\right)/2}\sum_{n=1}^{\infty}\frac{1}{n^{\left(d+1\right)/2}}K_{\left(d+1\right)/2}\left(n\beta M\right)\left[e^{in\theta}+e^{-in\theta}\right].\end{equation}
Applying our summation formula to $V_{B}$, we have\begin{equation}
\sum_{k=0}^{m-1}V_{B}\left(\beta,\theta+\frac{2\pi k}{m},M\right)=m^{-d}V_{B}\left(\beta,m\theta,mM\right)\end{equation}
from which a series of useful formulas for evaluating the effective potential
in phases with various $Z(N)$ symmetries can be derived.

The effective potential of an adjoint $SU(N)$ massive boson in the confinining
phase is given by\begin{equation}
\sum_{j,k=1}^{N}V_{B}\left(\beta,\frac{2\pi\left(j-k\right)}{N},M\right)-V_{B}\left(\beta,0,M\right)\end{equation}
 where the last term subtracts out the singlet term that would occur
for $U(N)$. Applying the summation formula, this is simply\begin{equation}
N^{1-d}V_{B}\left(\beta,0,NM\right)-V_{B}\left(\beta,0,M\right).\end{equation}
In the case where $N$ is even and $Z(N)$ is spontaneously broken
to $Z(N/2)$, the effective potential is\begin{equation}
4\sum_{j,k=1}^{N/2}V_{B}\left(\beta,\frac{2\pi\left(j-k\right)}{N/2},M\right)-V_{B}\left(\beta,0,M\right)\end{equation}
which reduces to\begin{equation}
4\left(N/2\right)^{1-d}V_{B}\left(\beta,0,NM/2\right)-V_{B}\left(\beta,0,M\right).\end{equation}

Now consider massless gauge bosons with periodic boundary conditions
combined with $N_{f}$ adjoint Dirac fermions, also with periodic
boundary conditions. The effective potential of the confined phase is\begin{equation}
\left[N^{1-d}V_{B}\left(\beta,0,0\right)-V_{B}\left(\beta,0,0\right)\right]-2N_{f}\left[N^{1-d}V_{B}\left(\beta,0,NM\right)-V_{B}\left(\beta,0,M\right)\right]\end{equation}
and the effective potential of the phase where $Z(N)$ is spontaneously
broken to $Z(N/2)$ is\begin{equation}
\left[4\left(N/2\right)^{1-d}V_{B}\left(\beta,0,0\right)-V_{B}\left(\beta,0,0\right)\right]-2N_{f}\left[4\left(N/2\right)^{1-d}V_{B}\left(\beta,0,NM/2\right)-V_{B}\left(\beta,0,M\right)\right].\end{equation}
The two minima of the effective potential are equal when\begin{equation}
\left(2^{d+1}-1\right)V_{B}\left(\beta,0,0\right)=2N_{f}\left[2^{d+1}V_{B}\left(\beta,0,NM/2\right)-V_{B}\left(\beta,0,NM\right)\right].\end{equation}
From this we find that $\left(\beta M\right)_{c}$ decreases as $1/N$;
in particular, \begin{equation}
\left(\beta M\right)_{c}\simeq\frac{4.00398}{N}\end{equation}
for $d=4$ and $N_{f}=1$, consistent with \cite{Myers:2008ey}.
 Note that a solution exists only for $N_{f}>1/2$
and $\left(\beta m\right)_{c}$ generally decreases as $N_{f}$ increases.
This analysis can easily be extended to evaluate the effective potential when
$Z(N)$ is spontaneously broken to $Z(p)$, where $p$ is an arbitrary
factor of $N$. Such expressions are useful for cases such as $N=6$,
where all the phases have a $Z(p)$ symmetry, but do not apply for
general $N$, due to the occurrence of phases without $Z(p)$ invariance
\cite{Myers:2009df}.

\end{document}